\documentclass[aps,twocolumn,tightenlines]{revtex4}
\usepackage[dvipdf]{graphicx}
\usepackage{color}
\usepackage{amssymb,amsmath}

\begin{document}
\title{Test of the He-McKellar-Wilkens topological phase by atom interferometry. \\ Part I:
theoretical discussion}

\author{S.~Lepoutre, A.~Gauguet, M.~B\"uchner, and J.~Vigu\'e}
\address{ Laboratoire Collisions Agr\'egats R\'eactivit\'e -IRSAMC
\\Universit\'e de Toulouse-UPS and CNRS UMR 5589
 118, Route de Narbonne 31062 Toulouse Cedex, France
\\ e-mail:~{\tt jacques.vigue@irsamc.ups-tlse.fr}}

\date{\today}

\begin{abstract}

We have recently tested the topological phase predicted by He and
McKellar and by Wilkens: this phase appears when an electric dipole
propagates in a transverse magnetic field. In the present paper, we
first recall the physical origin of this phase and its relations to
the Aharononov-Bohm and Aharonov-Casher phases. We then explain
possible detection schemes and we briefly describe the lithium atom
interferometer we have used for this purpose. Finally, we analyze in
great detail the phase shifts induced by electric and magnetic fields
acting on such an interferometer, taking into account
experimental defects. The experiment and its results are described in
a companion paper.

\end{abstract}
\maketitle
\bigskip

\noindent {\bf Keywords} topological phases; Aharononov-Bohm;
Aharonov-Casher; He-McKellar-Wilkens; atom interferometry;  Stark
effect; Zeeman effect\\

\section{Introduction}

In 1993, X.G. He and B.H.J. McKellar \cite{HePRA93} predicted a new
topological phase when an electric dipole encircles a line of
magnetic monopoles. Magnetic monopoles being hypothetical
\cite{milton06}, this idea seemed purely speculative but, in 1994,
M.~Wilkens \cite{WilkensPRL94} proposed an experimental test with an
atom (or a molecule) polarized by an electric field interacting with
a feasible magnetic field. This topological phase is now called the
He-McKellar-Wilkens (HMW) phase and it is the third electromagnetic
topological phase, after the Aharonov-Bohm \cite{Aharonov59} and
Aharonov-Casher phases \cite{AharonovPRL84}.

We have recently made an experimental test of the HMW phase
\cite{LepoutrePRL12}. The present paper describes the theory of our
experiment, which analysis and results are given in a
companion paper \cite{LepoutreXXX} called here HMWII. Section \ref{theory}
explains the nature of a topological phase and recalls what
the Aharonov-Bohm phase is. We then discuss the Aharonov-Casher and HMW phases and the
connections between these three effects. In section \ref{detection},
we describe various possible ways of detecting the HMW phase and the
principle of our experiment. In section \ref{s2}, we calculate the
effects of phase dispersions on the fringe signal of an atom
interferometer. In sections \ref{s3} and \ref{s4}, we evaluate the
phases induced by electric and magnetic fields in a lithium atom
interferometer. In section \ref{s5}, we evaluate the Aharonov-Casher phase in our experiment and in section \ref{s5}, we summarize the various phase shifts present in our experiments, their magnitude, their velocity dispersion, their internal state dependence and their effect on fringe visibility.

\section{Electromagnetic topological phases: theory and previous experiments}
\label{theory}

Here, we explain the nature of a topological phase and
describe the Aharonov-Bohm,  Aharonov-Casher and HMW
phases, and their connections.

\subsection{Topological phases and Aharonov-Bohm effect}

Topological (or geometric) phases were introduced in their general
form in 1984 by M.V. Berry \cite{BerryPRS84} as phase factors
associated to adiabatic transport (for a review, see ref.
\cite{Shapere89}), and we will consider here only matter
waves. It is interesting to compare topological phases and dynamic
phases.
\begin{itemize}

\item A topological phase is a quantum effect without any other
modification of the particle propagation and it can be detected only
by interferometry. It is independent of the modulus of
the velocity but it changes sign with the direction of propagation.

\item A dynamic phase is induced by a classical force
acting on the particle and, at first order of perturbation theory,
it is proportional to the difference, between the two interferometer arms, of the potential energy from which
the force derives and it is also proportional to the interaction time.
Therefore, a dynamic phase scales like the inverse of
the particle velocity and is independent of the
direction of propagation. Moreover, the classical force can be
detected by other experiments such as the deflection of the particle
trajectory or by the modification of its time-of-flight.

\end{itemize}

The vectorial Aharonov-Bohm (AB) phase \cite{Aharonov59}, discovered
in 1959, appears when a charged particle propagates in an
electromagnetic time-independent vector potential. The proposed
experiment (see fig. 2 of ref. \cite{Aharonov59}) involved an
electron interferometer with its arms encircling an infinite
solenoid. The AB phase shift reads:

\begin{equation}
\varphi_{AB} =  \frac{q}{\hbar} \oint \mathbf{A}
\left(\mathbf{r}\right) d \mathbf{r} = \frac{q}{\hbar} \Phi_0
\label{PhiAB}
\end{equation}

\noindent where $q$ is the electron charge, $\mathbf{r}$ is the
electron position and the closed circuit follows the interferometer
arms, $\mathbf{A} \left(\mathbf{r} \right)$ is the vector potential
and $\Phi_0$ is the total magnetic flux through any surface lying on
the closed circuit (the same result is obtained if the solenoid is
replaced by an infinite line of magnetic dipoles). In the proposition
of Aharonov and Bohm, the magnetic field vanishes on the
interferometer arms and the particle does not feel any
force, nevertheless the AB phase does not vanish. A
controversy followed this surprising prediction
\cite{ErlichsonAJP70,OlariuRMP85} but the AB effect was observed as
soon as 1960 by R.G. Chambers \cite{ChambersPRL60} and, thanks to
progress in electron interferometry, all the striking characteristics
of the AB effect have been tested experimentally
\cite{TonomuraPRL86,Peshkin89}.

M.V. Berry interpreted the vectorial Aharonov-Bohm phase as a
geometric phase \cite{BerryPRS84}: the common use is to call
topological the AB phase and to call geometric a phase acquired
through adiabatic transport but there are no fundamental differences
between these two types of phase. The AB effect is the first member
of a family of three topological phases occurring in the propagation
of particles in time-independent electromagnetic potentials or
fields, the other members being the Aharonov-Casher (AC) phase and the
He-McKellar-Wilkens (HMW) phase.

\subsection{Theory of the Aharonov-Casher phase \label{SubSecACPhase}}

\begin{figure}[h!]
\begin{center}
\includegraphics[height = 4cm]{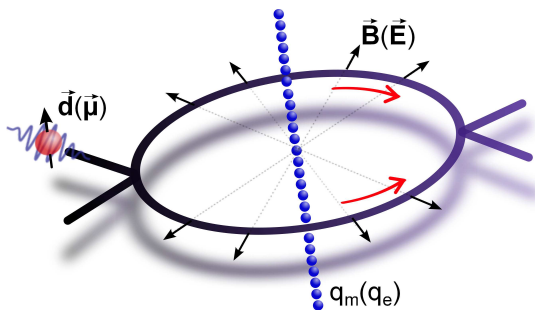}
\end{center}
\begin{center}
\caption{\label{figure1new} (color online). Connection between the HMW and AC phases
by electric-magnetic duality. The HMW phase arises when an electric
dipole moment $\mathbf{d}$ propagates in the radial magnetic field
created by a line of magnetic monopoles $q_m$ while the AC phase
(between parenthesis) appears when a magnetic dipole
$\boldsymbol{\mu}$ encircles a line of electric charges $q_e$.}
\end{center}
\end{figure}

An Aharonov-Bohm phase appears when a charged particle encircles an
infinite line of magnetic dipoles. By exchanging the roles of the
charged particle and of the magnetic dipole, Y. Aharonov and A.
Casher \cite{AharonovPRL84} predicted in 1984 a topological phase
when a magnetic dipole encircles an infinite line of electric charges
(see fig. \ref{figure1new}). This phase had already been predicted in 1982 by
J. Anandan \cite{AnandanPRL82}, with no insistence on its topological
character. The Aharonov-Casher (AC) phase is given by:

\begin{equation}
\varphi_{AC} = - \frac{1}{\hbar c^2} \oint \left[\mathbf{E} \left( \mathbf{r} \right) \times \boldsymbol{\mu} \right] \cdot d \mathbf{r}
\label{PhiAC}
\end{equation}

\noindent where $\boldsymbol{\mu}$ is the magnetic dipole and
$\mathbf{E}$ the electric field. As for the AB effect, the nature of
the AC effect was widely discussed
\cite{KleinPhysica86,BoyerPRA87,AharonovPRA88,ReznikPRD89,GoldhaberPRL89,AnandanPLA89,LiangPRL89,VaidmanAJP90,LiangMPLA90,HagenPRL90,HagenIJMP91,RubioNC91,MignaniJPA91,GoldhaberPLB91,ZeilingerPLA91,HePLB91,HePLB91Bis,HolsteinAJP91,LiangIJMP92,HanPLA92,SpavieriEPL92,LiangPLA93,RamseyPRA93,BalatskyPRL93,ChoiPRL93,ReznikPLB93,SpavieriNC94,LeePLA94,PeshkinPRL95,AlJaberNC95,LeeMPL96,FreemanEJP97,PeshkinFoP99,AharonovPRL00,SpavieriNC00,SwanssonJPA01,BoyerFoP02,HyllusPRL02,AharonovPRL02,DulatPRL12}.
In the non-relativistic limit (an excellent approximation for matter
wave interferometers if we except electron interferometers), we can
demonstrate eq. \ref{PhiAC}, starting from the Lagrangian of a
particle of mass $m$ and velocity $\mathbf{v}= \mathbf{\dot{r}}$
carrying a magnetic dipole $\boldsymbol{\mu}$ in an electric field
\cite{AharonovPRL84}:

\begin{equation}
L = \frac{1}{2} m \mathbf{v}^2 - \frac{1}{c^2} \mathbf{v} \cdot
\left( \mathbf{E} \left( \mathbf{r} \right) \times \boldsymbol{\mu}
\right) \label{LAC}
\end{equation}

\noindent The particle acceleration $\mathbf{\dot{v}} $ is given by
Lagrange equation \cite{AharonovPRA88}:

\begin{equation}
m \mathbf{\dot{v}} = \left( \boldsymbol{\mu} \cdot \nabla \right)
\left( \frac{\mathbf{E} \left(\mathbf{r} \right) \times
\mathbf{v}}{c^2} \right) \label{AccAC}
\end{equation}

\noindent In the configuration of ref. \cite{AharonovPRL84}, with a
straight homogeneously charged line, the right-hand term of eq.
\ref{AccAC} vanishes: no force acts on the particle.

A heuristic point of view introduced by A.G. Klein
\cite{KleinPhysica86} relates the AC phase results to the interaction
of the magnetic moment $\boldsymbol{\mu}$ with the motional magnetic
field $\mathbf{B}_{mot}\approx - \left(\mathbf{v} \times
\mathbf{E}\right) / c^2$ experienced by the particle in its rest
frame, and calculated at first order in $v/c$.
Substituting $d\mathbf{r} = \mathbf{v} dt$ into eq. \ref{PhiAC}
yields $\varphi_{AC} = \oint \left( \boldsymbol{\mu} \cdot
\mathbf{B}_{mot} \right) dt/ \hbar $, a result identical to the phase
due to the magnetic dipole interaction $-\boldsymbol{\mu} \cdot
\mathbf{B}_{mot}$. Eq. (\ref{LAC}) reads $L = m \mathbf{v}^2 / 2 +
L_{AC}$, where $L_{AC} = \boldsymbol{\mu} \cdot \mathbf{B}_{mot}$ is
the additional term due to the electric field. Although $-L_{AC}$
looks like a potential energy, it is not a potential energy for the
motion of the particle, because $m \mathbf{\dot{v}}$ given by
equation (\ref{AccAC}) is not equal to $\nabla \left(
\boldsymbol{\mu} \cdot \mathbf{B}_{mot} \right)$. Indeed, the use of
Newton's equation with the force $\nabla \left( \boldsymbol{\mu}
\cdot \mathbf{B}_{mot} \right)$ leads to incorrect results with
regards to the topological nature of the AC phase
\cite{BoyerPRA87,AharonovPRA88,GoldhaberPRL89}.

To deduce the AC phase from the Lagrangian (eq. (\ref{LAC})), we apply
Feynman's path integral  \cite{FeynmanRMP48} to matter-wave
interferometry \cite{CohenJDP94}. At first-order of perturbation
theory, the phase $\varphi_{AC}$ is given by the classical action
calculated along the unperturbed interferometer arms:

\begin{equation}
\varphi_{AC} = \frac{1}{\hbar} \oint \mathbf{p}_{AC} \cdot d \mathbf{r}
\label{PhiACAction}
\end{equation}

\noindent where $\mathbf{p}_{AC} = \partial L_{AC} / \partial
\mathbf{v} = - \mathbf{E} \left( \mathbf{r} \right) \times
\boldsymbol{\mu} / c^2$ is the modification of the particle momentum
by the electric field.

\subsection{Detection of the AC phase \label{SubSecDetAC}}

The AC phase was first detected by A. Cimmino \textit{et al.}
\cite{CimminoPRL89,KaiserPhysicaB88} using a neutron interferometer.
The neutron magnetic dipole is small and the AC phase was only $1.50$
mrad for $E \approx 30$ MV/m. Because of limited neutron flux, $35$
days were needed to get one measurement. Further tests
(proportionality to the electric field, independence with neutron
velocity) were not feasible.

A noticeable difference between the AB and AC phases is that the
particle must propagate in an electric field to get a non-zero AC
phase. This circumstance gives more freedom in the field
configurations and, in particular, the electric charge between the
interferometer arms may vanish \cite{AnandanPLA89,CasellaPRL90}. K.
Sangster \textit{et al.} \cite{SangsterPRL93,SangsterPRA95} used this possibility to perform a measurement of the
AC phase with a Ramsey interferometer \cite{RamseyPR50}: a molecular
beam, prepared in a coherent superposition of states with opposite
magnetic dipoles, propagates in an electric field perpendicular to
the beam velocity. The AC phase shift has opposite values for these
two states and the resulting phase difference is directly detected on
the fringe signal. The AC phase, measured with a few percent error
bar, was found in agreement with theory
\cite{SangsterPRL93,SangsterPRA95}; its proportionality to the
electric field and its velocity independence were both successfully
tested. Several other measurements of the AC phase have been
performed, always with Ramsey interferometers
\cite{ZeiskeAPB95,GorlitzPRA95,YanagimachiPRA02}. The AC effect has
also been observed in interference of vortices in a
Josephson-junction array \cite{ElionPRL93}.

\subsection{The He-McKellar-Wilkens phase}

In 1993,  X.G.~He and B.H.J.~McKellar \cite{HePRA93} applied Maxwell
duality to the AC phase, thus predicting a new topological phase when
a particle with an electric dipole $\mathbf{d}$ encircles an infinite
line of magnetic monopoles (see fig. \ref{figure1new}). Because of
the hypothetical character of magnetic monopoles \cite{milton06},
this paper did not suggest any test but M.~Wilkens
\cite{WilkensPRL94} proposed an experiment, with an electric dipole
produced by the polarization of an atom or a molecule, interacting with
a magnetic field $\mathbf{B}$ guided by ferromagnetic materials. The
general expression of the HMW phase is:

\begin{equation}
\varphi_{HMW} = \frac{1}{\hbar} \oint \left( \mathbf{B} \left(\mathbf{r}
\right) \times \mathbf{d} \right) \cdot d \mathbf{r} \label{PhiHMW}
\end{equation}

\noindent Fig. \ref{figure1new} is inspired by the work of J.P.
Dowling \textit{et al.} \cite{DowlingPRL99} who gave an overview of
the electromagnetic topological phases. Maxwell duality applied to
the AB phase leads to a fourth topological phase for a magnetic
monopole encircling a line of electric dipoles: this phase will
remain hypothetical as long as magnetic monopoles.

The AB phase involves a particle carrying an electric charge and the
AC and HMW phases involve particles carrying magnetic and electric
dipoles: it is natural to predict topological phases for particles
carrying higher-order electromagnetic multipoles,  in interaction
with electric or magnetic fields of the convenient symmetry. A
calculation for the case of electric or magnetic quadrupoles was made
by C.-C. Chen \cite{ChenPRA95} who states that, with quadrupoles of
the order of one atomic unit, the detection of these new topological
phases "would require an unrealistically huge electromagnetic field".
As a consequence, these higher-order phases appear to be out of reach
and the HMW phase was the last undetected topological phase of
electromagnetic origin.

\subsection{Some properties of the HMW effect and the associated particle dynamics}

In complete analogy with the AC effect, the HMW effect can be
interpreted as due to the interaction of the electric dipole
$\mathbf{d}$ with the motional electric field $\mathbf{E}_{mot}
\approx \mathbf{v} \times \mathbf{B}$, at the lowest order in $v/c$.
Equation (\ref{PhiHMW}) can be rewritten:

\begin{equation}
\varphi_{HMW} = \frac{1}{\hbar} \oint \mathbf{d} \cdot \mathbf{E}_{mot} dt \label{PhiHMWa}
\end{equation}

\begin{figure}[h!]
\begin{center}
\includegraphics[width=8 cm]{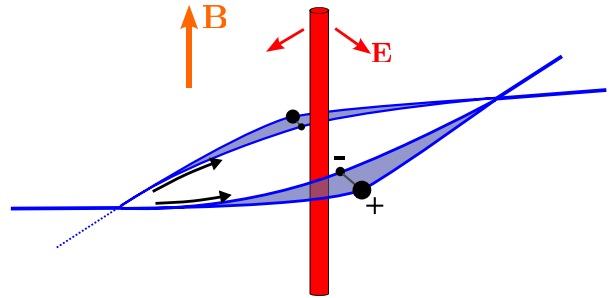}
\end{center}
\begin{center}
\caption{\label{figure2new} (color online). Connection between the HMW and the AB
phases following Wei \textit{et al.} \cite{WeiPRL95}.
The interferometer arms (blue full lines) encircle an infinite
charged wire (red vertical cylinder) which produces a radial electric
field $\mathbf{E}$ and induces an electric dipole
$\mathbf{d}$ represented by a positive charge ($+$, large bullet) and
a negative charge ($-$, small bullet). Each charge undergoes the AB
effect in the uniform magnetic field $\mathbf{B}$. The HMW phase is
equal to the sum of the two AB phases and it is proportional to the
magnetic flux through the (blue) shaded surface.}
\end{center}
\end{figure}

\noindent A remark first done by Wei \textit{et al.} \cite{WeiPRL95}
suggests a strong link between the AB and the HMW phases.
Consider the particular field configuration illustrated
by fig. \ref{figure2new}, where the electric dipole which undergoes
the HMW phase shift is induced by an external electric field (more
details are given in part \ref{SubSecDetHMW}). If the dipole is
described by two particles with charges $\pm q$ at positions
$\mathbf{r}_\pm$, with $\mathbf{d} = q \left(\mathbf{r}_+ -
\mathbf{r}_-\right)$, the HMW phase is equal to the algebraic sum of the AB
phases for the two particles.

\noindent

The HMW effect and its connection with the AB and AC effects has been
the subject of many theoretical works
\cite{LiuCPL95,YiPRB95,SpavieriNC96,HagenPRL96,WeiPRL96,SpavieriPRL98,WilkensPRL98,LeonhardtEPL98,AudretschPLA98,AudretschPRA99,LeonhardtPLA99,AudretschPLA99,SpavieriPRA99,SpavieriPRL99,TkachukPRA00,AnandanPRL00,LeePRA00,LeePRA01,SpavieriPLA03,FurtadoPRA04,IvezicPRL07}.
Let us summarize the main results concerning the dynamics of an
electric dipole in a magnetic field. The electric dipole moment
$\mathbf{d} = q \mathbf{r}_0$ is described by two charges with
$\mathbf{r}_0 = \mathbf{r}_+ - \mathbf{r}_-$ and its internal
dynamics is described by an interaction energy $U \left( r_0
\right)$, function of the distance $r_0$ between the charges. The
compound particle (mass $M$, center of mass  $\mathbf{r}$, velocity
$\mathbf{v} = \dot{\mathbf{r}}$) interacts with an external
electromagnetic field described by its potential $\left(\mathbf{A}
\left( \mathbf{r}, t \right), V \left( \mathbf{r},t \right) \right)$,
with the electric field $\mathbf{E} \left(\mathbf{r}, t \right) = -
\nabla V - \partial \mathbf{A} / \partial t$ and the magnetic field
$\mathbf{B} \left( \mathbf{r}, t \right) = \nabla \times \mathbf{A}$.
The standard Lagrangian for the system, expressed in the dipole
approximation, is \cite{SpavieriPRA99}:

\begin{eqnarray} \label{LEBStandard}
L &=& \frac{1}{2} M {\dot{\mathbf{r}}}^2 + \frac{1}{2} \mu
{\dot{\mathbf{r}}_0}^2 + \dot{\mathbf{r}} \cdot \left[ \left(
\mathbf{d} \cdot \nabla \right) \mathbf{A} \left( \mathbf{r},
t\right) \right] \\ \nonumber & &\quad + \dot{\mathbf{d}} \cdot
\mathbf{A} \left( \mathbf{r}, t\right) - \left( \mathbf{d} \cdot
\nabla \right) V \left( \mathbf{r}, t \right) - U \left(r_0 \right)
\end{eqnarray}

\noindent where $M$ is the total mass of the compound particle and
$\mu$ is the reduced mass of the two particles. Expanding the total
derivative $d/dt = \partial / \partial t + \left( {\dot{\mathbf{r}}}
\cdot \nabla \right)$, it can be rewritten:

\begin{eqnarray}\label{LEBW}
L &=& \frac{1}{2} M {\dot{\mathbf{r}}}^2 + \frac{1}{2}\mu
{\dot{\mathbf{r}}_0}^2 + L_W - U \left(r_0 \right) + \frac{d}{d t}
\left( \mathbf{d} \cdot \mathbf{A} \right)\\ \nonumber && \text{with:
} L_W = \mathbf{d} \cdot \left( \mathbf{E} + \mathbf{v} \times
\mathbf{B} \right)
\end{eqnarray}

\noindent $L_{W}$ is the term introduced by M. Wilkens
\cite{WilkensPRL94} to describe the interaction of the dipole with
the field. In his calculation, the total derivative term $d\left(
\mathbf{d} \cdot \mathbf{A} \right)/dt$ was omitted. Because this
total derivative is a single valued function of the dynamical
variables and of time, the standard Lagrangian and the Lagrangian
proposed by Wilkens are strictly equivalent
\cite{SpavieriPRL98,WilkensPRL98}. With the Lagrangian used by
Wilkens, the canonical momenta are given by:

\begin{eqnarray}
\mathbf{p} &=& \frac{\partial L}{\partial \dot{\mathbf{r}}} = M
\mathbf{v} + \mathbf{B} \times \mathbf{d}\\
\mathbf{p}_0 &=& \frac{\partial L}{\partial {\dot{\mathbf{r}}}_0} =
\mu {\dot{\mathbf{r}}}_0
\end{eqnarray}

\noindent As for the AC effect, the extra-contribution
$\mathbf{p}_{HMW}= \mathbf{B} \times \mathbf{d}$ to the momentum
$\mathbf{p}$ yields the HMW phase:

\begin{equation}
\varphi_{HMW} = \frac{1}{\hbar} \oint \mathbf{p}_{HMW} \cdot
d\mathbf{r} \label{PhiHMWc}
\end{equation}

The Lagrange equations yield the dynamics of the particle in the
laboratory frame and its internal dynamics:

\begin{eqnarray}
M\ddot{\mathbf{r}} &=& \dot{\mathbf{d}} \times \mathbf{B} +
\left(\mathbf{d} \cdot \nabla \right) \left[ \mathbf{E} + \mathbf{v}
\times \mathbf{B} \right] \label{DynPart}\\
\mu \ddot{\mathbf{r}}_0 &=& q \left(\mathbf{E} + \mathbf{v} \times
\mathbf{B} \right) - \frac{\partial U}{\partial \mathbf{r}_0}
\label{DynDipole}
\end{eqnarray}

In the original configuration  with an infinite line of magnetic
monopoles line  (see Fig. \ref{figure1new}), the force on the
particle vanishes. Here is a brief summary of the explanation given
by M. Wilkens \cite{WilkensPRL94}. The dipole dynamics is limited to
rotation, with the dipole initially parallel to the line of
monopoles, while the particle propagates in a plane perpendicular to
this line. The torque exerted on the dipole, $\mathbf{d} \times
\left( \mathbf{E} + \mathbf{v} \times \mathbf{B} \right)$ vanishes
and we may drop the term $\dot{\mathbf{d}} \times \mathbf{B}$ from
Eq. (\eqref{DynPart}). If the fields $\mathbf{E}$ and $\mathbf{B}$
are invariant by a translation along the direction of the dipole, the
force vanishes.

A closer look is needed in the case of an induced dipole. In this
case, it is a good approximation to consider that the variations of
the external fields in the frame moving with the atom are infinitely
slow and that the dynamics of $\mathbf{r_0}$ is adiabatic, so that
the atom exhibits the dipole $\mathbf{d} = 4 \pi \varepsilon_0 \alpha
\left(\mathbf{E} + \mathbf{v} \times \mathbf{B}\right)$, where
$\alpha$ is the polarizability. With this approximation, one obtains
the Lagrangian proposed by Wei \textsl{et al} \cite{WeiPRL95}:

\begin{equation}
L = \frac{1}{2} M {\dot{\mathbf{r}}}^2 + 2 \pi \varepsilon_0 \alpha {\left(\mathbf{E} + \mathbf{v} \times \mathbf{B} \right)}^2
\label{LWei}
\end{equation}

\noindent From this Lagrangian, it is easy to deduce the force on the
atom and we consider here only three terms which involve
the presence of $\mathbf{E}$ and $\mathbf{B}$ simultaneously:

\begin{eqnarray}
\mathbf{F}_1 &=& 4 \pi \varepsilon_0 \alpha \left( \mathbf{\dot{E}} \times \mathbf{B}  \right)\nonumber \\
\mathbf{F}_2 &=& 4 \pi \varepsilon_0 \alpha \left(\mathbf{v} \times \mathbf{B}\right)\cdot \nabla \mathbf{E} \nonumber \\
\mathbf{F}_3 &=& 4 \pi \varepsilon_0 \alpha \mathbf{E} \cdot \nabla \left(\mathbf{v} \times \mathbf{B}\right)
\label{LWei2}
\end{eqnarray}

\noindent We assume that both fields are static i.e. $\partial
\mathbf{E}/\partial t= \mathbf{0} = \partial \mathbf{B}/\partial t$.
$\mathbf{F}_1$ is non-zero in the regions where the electric field is
inhomogeneous because, in the atom frame, $\mathbf{\dot{E}} =
\left(\mathbf{v} \cdot \nabla \right) \mathbf{E}$. When the electric
field varies, the dipole varies too, which induces a current, and
$\mathbf{F}_1$ is the associated Lorentz force. If we take the
$\mathbf{z}$ axis along the velocity $\mathbf{v}= v \mathbf{z}$ and
the $\mathbf{y}$ axis along the magnetic field $\mathbf{B} = B
\mathbf{y}$, we are interested only in the $z$-components of the
forces because they are the only ones which can change the velocity:

\begin{eqnarray}
F_{1,z} &=& 4 \pi \varepsilon_0 \alpha v  B \frac{\partial E_x}{\partial z}   \nonumber \\
F_{2,z} &=& - 4 \pi \varepsilon_0 \alpha v  B \frac{\partial E_z}{\partial x} \nonumber \\
F_{3,z} &=& 0
\label{LWei3}
\end{eqnarray}

As $\nabla \times \mathbf{E} = - \partial \mathbf{B}/\partial t= 0$, then $\partial E_x/\partial z = \partial E_z/\partial x$ and   $ F_{1,z} + F_{2,z} =0$: in the Wilkens-Wei configuration \cite{WilkensPRL94,WeiPRL95}, the HMW phase is a topological phase.

\section{Toward a detection of the HMW phase }
\label{detection}

In this section, we describe various proposals for the
detection of the HMW phase, and we  explain the choices
we have made for our experiment.

\subsection{Possible detection schemes}
\label{SubSecDetHMW}

\begin{figure}[h!]
\begin{center}
\includegraphics[width = 6 cm]{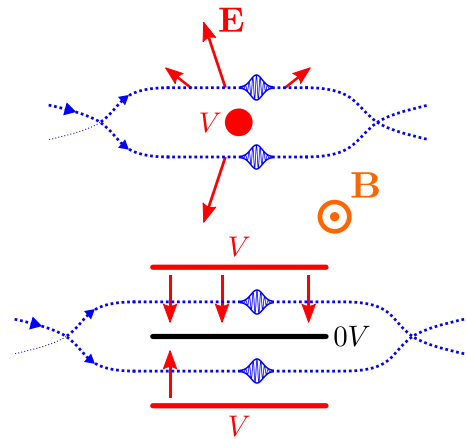}
\end{center}
\begin{center}
\caption{\label{figure3new} (color online). Geometry of the electric field for the
detection of the HMW phase, following the proposal of Wei \textit{et
al.} \cite{WeiPRL95}. The atom (blue dots) propagates in the
interferometer plane, with the homogeneous magnetic field
$\mathbf{B}$ perpendicular to this plane. The conductors are shown
with their potential (red or black if grounded) and the electric
field vector $\mathbf{E}\left(\mathbf{r}\right)$ is represented by
red arrows for some sample positions along the interferometer arms.
Upper panel (similar to fig. \ref{figure2new}): the
original proposal with a charged wire which produces a
strongly inhomogeneous electric field. Lower panel: our geometry with
homogeneous electric fields produced by plane capacitors.}
\end{center}
\end{figure}

The detection of the HMW phase is difficult for two reasons: we need
a particle with an electric dipole and we must replace the radial
magnetic field of a line of magnetic monopoles by some other field
configuration. Here are the experimental schemes proposed for this
detection:

\begin{itemize}

\item M. Wilkens \cite{WilkensPRL94} proposed to polarize atoms (or
molecules) by an electric field $\mathbf{E}$, thus inducing a dipole
$\mathbf{d} = 4\pi \varepsilon_0 \alpha \mathbf{E}$,  and to apply
different magnetic fields on the two interferometer arms thanks to a
pierced sheet of ferromagnetic material. The use of such a sheet
appears to be very difficult because of the small distance between
interferometer arms and of the associated perturbation of the
electric field. However, this proposal opened the way toward
experiments.

\item H. Wei \textit{et al.} \cite{WeiPRL95} proposed to introduce a
charged wire between the arms of an atom interferometer, thus
inducing opposite dipoles on the two interferometer
arms, and to use a common homogeneous magnetic field to
induce the HMW phase. Figures \ref{figure2new} and
\ref{figure3new} illustrates this scheme, and figure \ref{figure3new}
depicts our own configuration which is directly inspired by this
proposal.

\item  H. Wei \textit{et al.} \cite{WeiPRL95} also predicted a
persistent current in a polarizable superfluid (see also
\cite{YiPRB95}). Following this idea, Y. Sato and R. Packard
\cite{SatoJOP09} have proposed to detect the HMW phase with a
superfluid helium interferometer.

\end{itemize}

It would be very convenient to use a Ramsey interferometer to
detect the HMW phase, as done for most of the AC phase measurements
\cite{SangsterPRL93,SangsterPRA95,ZeiskeAPB95,GorlitzPRA95,YanagimachiPRA02}.
Such an interferometer requires a coherent superposition of states
with opposite electric dipole moments
\cite{SpavieriPRL99,DowlingPRL99}, which seems feasible with molecules
or with Rydberg atoms, because they have quasi-degenerate states of
opposite parity \cite{DowlingPRL99}, but not with ground state atoms.
As a consequence, Ramsey interferometry with ground state atoms
cannot be used for the detection of the HMW phase. Instead, the HMW phase will be given by the difference of successive phase
measurements, a technique more sensitive to systematic
effects than Ramsey interferometry.

\subsection{Principle of our experiment}
\label{s1}

\begin{figure}[t]
\begin{center}
\includegraphics[width= 8 cm]{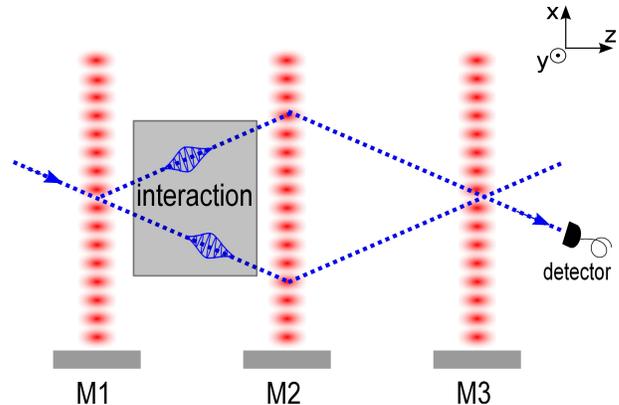}
\caption{(color online). Schematic top-view of our atom interferometer: the HMW
interaction region is placed just before the second laser standing
wave, at the place where the arm separation is largest. It is thus
possible to introduce a septum between the two interferometer arms
without any perturbation of the interferometer signal.
\label{figure4new}}
\end{center}
\end{figure}

To detect the HMW phase, we have built an experiment
\cite{LepoutrePhD,LepoutrePRL12} with our atom interferometer
\cite{MiffreEPJD05,MiffrePhD}  (see fig. \ref{figure4new}). A highly
collimated supersonic beam of lithium seeded in argon, with a mean
lithium velocity $v_m \approx 1065$ m/s, crosses three quasi-resonant
laser standing waves which diffract the atoms in the Bragg regime.
With first order Bragg diffraction which produces only two diffracted
beams (orders $p=0$ and either $p=+1$ or $p=-1$), we get in this way
an almost perfect Mach-Zehnder interferometer. A slit selects one of
the two output beams carrying complementary interference signals and
the intensity $I$ of this beam, measured by a surface
ionization detector, is the output signal of the interferometer:

\begin{equation}
\label{a0}
I=I_0 \left[1 + \mathcal{V}\cos\left(\varphi_d+\varphi_p\right)\right]
\end{equation}

\noindent $I_0$ is the mean intensity, $\mathcal{V}$ is the fringe
visibility and $\varphi_p$ is the phase due to various perturbations.
The phase $\varphi_d $, due to laser diffraction, is a function of
the positions $x_i$ of the three standing wave mirrors
M$_i$: $\varphi_d =2k_L(x_1-2x_2+x_3)$, where $k_L$ is the laser
wavevector. The choice of the laser frequency, very close to the
first resonance transition of lithium \cite{MiffreEPJD05}, and the
natural abundance of $^7$Li ($92.5$\%) make that the signal is purely
due to this isotope \cite{MiffreEPJD05,JacqueyEPL07}.

To observe a non-zero HMW phase, the atom must propagate in crossed
electric and magnetic fields transverse to its velocity and the
fields on the two interferometer arms must be different. Near the
second laser standing wave, the two arms are separated by a distance
close to $100$ $\mu$m, sufficient to insert a septum between the two
arms. A septum can be used to produce different magnetic fields by
circulating a current in the septum \cite{SchmiedmayerJPII94} or
different electric fields with two capacitors sharing the septum as a
common electrode \cite{EkstromPRA95}. The difference of magnetic
fields achieved in ref. \cite{SchmiedmayerJPII94} was quite small,
near $10^{-5}$ T, limited by the current in the septum, while the
second arrangement \cite{EkstromPRA95} has produced intense electric
fields, of the order of $1$ MV/m. We have chosen the second
arrangement with opposite electric fields on the two interferometer
arms and a common magnetic field: in addition to the HMW phase, this
arrangement produces several other phases discussed in sections \ref{s3} and \ref{s4}. This setup is very close to the idea
of Wei \textit{et al.} \cite{WeiPRL95} but the charged wire is
replaced by a septum, which improves considerably the electric field
homogeneity.

\section{Effect of a dispersion of the phase on the interferometer signal}
\label{s2}

Any dispersion of the phase $\varphi= \varphi_d+\varphi_p$ reduces
the fringe visibility ${\mathcal{V}}$ and a good visibility is
necessary for accurate phase measurements. In this part, we study the
origins of phase dispersions and the associated systematic effects.

\subsection{Origins of phase dispersion}
\label{s21}

The interferometer phase is dispersed because of its dependence with
the atom velocity, with the atom trajectory and with the atom
internal state.

The diffraction phase $\varphi_d$ is independent of the atom velocity
$v$ but the perturbation phase $\varphi_p$ is a priori a function of $v$. A dynamic phase
due to a perturbation applied to one arm is proportional to
$1/v$. If the same perturbation is applied to both arms,
the phase shift vanishes if the perturbation is
homogeneous and is proportional to $1/v^2$ in the presence of a
perturbation gradient, with an extra $1/v$-factor due to the distance
between the interferometer arms which is approximately proportional to $1/v$.
The topological AC and HMW phases are independent of the velocity.
Finally, inertial phase shifts are proportional to $1/v$ (Sagnac
effect) and to $1/v^2$ (homogeneous gravitational field): in our
experiment, there is a small Sagnac phase due to Earth rotation \cite{JacqueyPRA08} but the phase due to the gravitational
field vanishes because the interferometer is an horizontal plane.

In our experiment, the magnetic field is slightly inhomogeneous and
the electric fields have slightly different modulus on the two
interferometer arms. Atom diffraction is in the horizontal plane,
which means that the interferometer signal is sensitive to the
difference of the propagation phases on the two arms at the same
altitude $y$. The resulting phase shifts are functions
of the $y$-coordinate because of the spatial dependence of the
fields.

The diffraction phase shift $\varphi_d$ is also a function of the
$y$-coordinate, if the laser standing wave mirrors M$_i$ are not
perfectly aligned (for an analysis, see ref.
\cite{ChampenoisEPJD99,ChampenoisPhD}). The final alignment of these
mirrors is done by optimizing the fringe visibility
\cite{MiffreEPJD05} and this procedure is not sensitive to a small
residual $y$-dependence of $\varphi_d$.

The Zeeman phase is a function of the hyperfine-Zeeman $F,m_F$
sublevel; this phase, which may be large, varies rapidly with $F,m_F$ (see
section \ref{s4}). The interferometer signal is the sum of the
contributions of these 8 sublevels: in the absence of optical pumping,
the sublevels are equally populated in the incident atomic beam,
but the interferometer transmission is a function of the hyperfine
level $F$. As a consequence, the 8 sublevels may have
different populations in the detected signal: this question is
discussed in Appendix A.

\subsection{Effect of the velocity dependence of the phase-shifts}
\label{s22}

The normalized velocity distribution of a supersonic beam is given
by:

\begin{eqnarray}
\label{a2}
P(v)  = \frac{S_{\|}}{v_m \sqrt{\pi}}
\exp\left[-\left(\frac{\left(v-v_m\right)S_{\|}}{v_m}\right)^2\right]
\end{eqnarray}

\noindent $v_m$ is the mean velocity, $S_{\|}$ is the parallel speed
ratio. A $v^3$ pre-factor, usually included \cite{HaberlandRSI85},
has been omitted for two reasons: - when $S_{\|}$ is large, this
pre-factor has small effects; - because of the use of Bragg diffraction, the interferometer transmission is a
function of the velocity and this effect modifies the velocity distribution.
We consider a perturbation phase $\varphi_p(v) \propto 1/v^n$ so that we can write $\varphi_p(v)= \varphi_p(v_m) (v_m/v)^n$. The interferometer signal is the velocity-average of eq. (\ref{a0}):

\begin{equation}
\label{a3} I = I_0 \int dv P(v)  \left[1 + {\mathcal{V}} \cos\left(
\varphi_d + \varphi_p(v_m)  \left(\frac{v_m}{v}\right)^n \right) \right]
\end{equation}

\noindent If the ratio $\varphi_p(v_m)/S_{\|}$
is not too large, it is a good approximation to expand $v_m/v$ up to
the second order in powers of  $(v-v_m)/v_m$ and the integral can be
calculated analytically \cite{MiffreEPJD06,MiffrePhD}. The phase
shift differs from $\varphi_p(v_m)$ by a term linear in
$\varphi_p(v_m)/S_{\|}^2$ because of the non-linear dependence of
$\varphi_p$ with $v$ and the visibility decreases rapidly when
$\varphi_p(v_m) \approx S_{\|}/n $, with a quasi-Gaussian dependence.

\subsection{Calculation of the effect of a narrow distribution of phase shift}
\label{s23}

 Eq. (\ref{a3}) uses analytical expressions of
$\varphi(v)$ and of $P(v)$. For other types of phase dispersion,
this information is not generally available. For instance, for the
dependence of the phase with the atom trajectory, we may assume that
the phase is a function $\varphi(y)$ of a continuous variable $y$
with a normalized probability $P(y)$ and we must average
eq. (\ref{a0}):

\begin{equation}
\label{a4} \left<I\right> = I_0 \int dy P(y)  \left[1 + {\mathcal{V}}
\cos\left( \varphi(y) \right) \right]
\end{equation}

\noindent $\left< ...\right>$  denotes the average over
$y$ with the weight $P(y)$. We assume that the visibility
${\mathcal{V}}$ is independent of $y$ because the fringe visibility
has a very low sensitivity to the diffraction amplitudes
\cite{MiffreEPJD05}. We introduce:

\begin{eqnarray}
\label{a5} \left< \varphi \right> = \int dy P(y) \varphi(y) \\ \nonumber
\delta \varphi = \varphi(y) - \left<\varphi\right>
\end{eqnarray}

\noindent Obviously  $\left< \delta \varphi \right>=0$. Assuming that
$\delta \varphi$ is small, we expand $\sin \left(\delta
\varphi\right)$ and $\cos\left(\delta \varphi\right)$ up to third
order in $\delta\varphi$ (these expansions are of reasonable accuracy
even if $\left|\delta \varphi\right| \approx 1$  rad).
Once averaged over $y$, eq. (\ref{a4}) is similar to eq. (\ref{a0})
with a modified visibility $\mathcal{V}_m$ and a modified phase
$\varphi_m$:

\begin{eqnarray}
\label{a7}
\mathcal{V}_m/\mathcal{V}_0 &=&  1 -\left<\left(\delta \varphi\right)^2/2\right>\nonumber \\
\varphi_m &=& \left<\varphi\right> - \left<\left(\delta \varphi\right)^3/6\right>
\end{eqnarray}

\noindent The reduced visibility $ \mathcal{V}_r= \mathcal{V}_m /
\mathcal{V}_0$ carries interesting information when two perturbations $a$ and $b$
inducing the phases $\varphi_a$ and $\varphi_b$ are  simultaneously
applied:

\begin{eqnarray}
\label{a8} \mathcal{V}_{r,a+b}=
\frac{\mathcal{V}_{m,a+b}}{\mathcal{V}_0} &=&  1
-\frac{\left<\left(\delta \varphi_a + \delta
\varphi_b\right)^2/2\right>}{2} \nonumber \\
&\approx& \mathcal{V}_{r,a}\mathcal{V}_{r,b} \left[ 1 -\left<\delta
\varphi_a \delta \varphi_b\right>\right]
\end{eqnarray}

\noindent By measuring three reduced visibility $\mathcal{V}_{r,a}$,
$\mathcal{V}_{r,b}$ and  $\mathcal{V}_{r,a+b}$, we have access to the
correlation $\left<\delta \varphi_a  \delta \varphi_b\right>$ of the
dispersions of the two phases. The phase shift $\varphi_m$ due to the
perturbation is not equal to the mean phase $\left<\varphi\right>$
because, even if, by definition, $\left< \delta \varphi \right>=0$,
$\left<\left(\delta \varphi\right)^3\right>$ is usually not equal to
$0$. Moreover, if two perturbations $a$ and $b$ are simultaneously
applied, the phase shifts are not additive, because of the
cross-terms $\left<\delta \varphi_a^2  \delta \varphi_b\right>$ and
$\left<\delta \varphi_a  \delta \varphi_b^2\right>$.

\subsection{Discrete average over Zeeman-hyperfine sublevels}
\label{s24}

The signal is given by:

\begin{eqnarray}
\label{a10} I& =& I_0 \sum_{j} P_{j}  \left[1 +
\mathcal{V}_{j}\cos\left(\left<\varphi_{j}\right>\right)\right]
\end{eqnarray}

\noindent where  the signal due to the sublevel $j$ is characterized
by a normalized population  $P_{j}$ ($\sum_{j} P_{j}=1$), a
visibility $\mathcal{V}_{j}$ and a phase $\varphi_{j}$. The
visibility $\mathcal{V}_{j}$ varies with the sublevel because the
reduction of visibility given by eq. (\ref{a7}) is a function of the
sublevel. The $-\left<\left(\delta \varphi_j\right)^3/6\right>$ term,
omitted in eq. (\ref{a10}), will be taken into account in the complete
calculation. For the contribution of sublevel $j$ to the signal, we
define a complex fringe visibility $\underline{\mathcal{V}}_j $ given
by:

\begin{eqnarray}
\label{a11} \underline{\mathcal{V}}_j = \mathcal{V}_j \exp\left(i
\left<\varphi_j\right>\right)
\end{eqnarray}

\noindent The complex visibility for the total signal is given by:
\begin{eqnarray}
\label{a12}
\underline{\mathcal{V}} =& \sum_j P_{j} \underline{\mathcal{V}}_j
\end{eqnarray}

\noindent This is a Fresnel construction from which we deduce the
modified fringe visibility $\mathcal{V}_m$ and phase  $\varphi_m$:

\begin{eqnarray}
\label{a13} \mathcal{V}_m & =& \sqrt{ \left[\sum P_j
\mathcal{V}_j\cos\left<\varphi_j\right>\right]^2 + \left[\sum P_j
\mathcal{V}_j \sin\left<\varphi_j\right>\right]^2} \nonumber \\
\tan \varphi_m &=& \left(\sum  P_j \mathcal{V}_j
\sin\left<\varphi_j\right>\right)/\left(\sum  P_j \mathcal{V}_j
\cos\left<\varphi_j\right>\right)
\end{eqnarray}

\noindent When the phases $\left<\varphi_j\right>$ are very close to
their mean, the resulting phase $\varphi_m$ is their weighted
average, but the weights are the products $P_j  \mathcal{V}_j$ and
not the populations $P_{j}$. This result has an important
consequence: when a perturbation modifies the visibility
$\mathcal{V}_j$, the modified phase $\varphi_m$ is not a simple
average of $\left<\varphi_j\right>$. In this case too, even without
the non-linear term $\left<\delta \varphi^3\right>/6$, the phase
shifts resulting from two perturbations are not additive, because the
weights $P_j  \mathcal{V}_j$ are different in the three cases : application of perturbation $a$, application of perturbation $b$ and simultaneous application of both perturbations.

\section{Effects of the electric field on the interferometer signals}
\label{s3}

An electric field induces a large phase due to Stark effect and a
small one due to Aharonov-Casher effect \cite{AharonovPRL84}.
 Because of its dependence on the magnetic dipole
moment, the AC phase appears as a modification of the
Zeeman effect and we will discuss it after the Zeeman phase in
section \ref{s5}.

\subsection{Effective Stark Hamiltonian}
\label{s31}

If we neglect hyperfine structure, an electric field $\mathbf{E}$
induces only a global displacement of lithium $^2$S$_{1/2}$ ground
state described by the Stark Hamiltonian $H_{S}$ :

\begin{equation}\label{hs2}
H_S = - 2\pi\varepsilon_0 \alpha \mathbf{E}^2
\end{equation}

\noindent $\alpha$ is the electric polarizability, $\alpha = (24.34
\pm 0.16)\times 10^{-30}$ m$^3$ \cite{MiffreEPJD06,JacqueyPRA08}.
Theoretical values \cite{PuchalskiPRA12} are considerably more
accurate and in good agrement with this experimental value. For our
largest field  $E_{max} \approx 0.8 $ MV/m, the Stark energy is
$E_{S} \approx 10^{-27}$ J  while the atom kinetic energy is $K=
mv_m^2/2 \approx 7 \times 10^{-21} $ J. With $E_S/K$ smaller than
$2\times 10^{-7}$, a first order perturbation calculation of the
Stark phase is fully justified:

\begin{equation}\label{hs2a}
\varphi_{S}=2\pi\varepsilon_0 \alpha \oint \mathbf{E}^2 dt/\hbar
\end{equation}

\noindent If the field  $E_{max} \approx 0.8 $ MV/m was applied on
one interferometer arm only, the Stark phase would be large,
$\varphi_{S}\approx 300$ rad. In the experiments devoted to the
detection of the HMW phase, opposite electric fields are applied on
the two interferometer arms, resulting in a very small
detected Stark phase shift.

Because of its $3/2$ nuclear spin, $^7$Li has $8$ hyperfine-Zeeman
sublevels. The Stark shift is only approximately independent of
the sublevel but this dependence is very weak. This question is very
important for atomic clocks and it has been studied theoretically by
Sandars \cite{SandarsPPS67} and Ulzega \textit{et al.}
\cite{UlzegaEPL06}: the results are in good agreement with
experiments for the cesium clock \cite{SimonPRA98,OspelkausPRA03}.
For $^7$Li, only the energy shift difference $\Delta \textrm{E}_S$ of
the $F=1$, $m_F=0$ and $F=2$, $m_F=0$ sublevels has been measured
\cite{MowatPRA72},  $ \Delta \textrm{E}_S/h =  - 0.061(2) \times
10^{-10} E^2 $ Hz  with $E$ in V/m. This measurement is in good
agreement with theoretical values \cite{KaldorJPB73,PuchalskiPRA12}.
The ratio of this differential shift to the mean energy shift is
$\Delta \textrm{E}_S/ \textrm{E}_S \approx 3 \times 10^{-6}$ and we
may deduce that the $F,m_F$-dependence of the Stark phase is
negligible in our experiment.

\subsection{Stark phase-shift of an ideal experiment}
\label{s32}

We first assume defect-free capacitors, with plane parallel
electrodes. We use the same notations as in ref. \cite{MiffreEPJD06}:
electrode spacing $h_i$ and length between the guard electrodes
$2a_i$. The electric field $\mathbf{E}_i(z)$ is easily calculated
\cite{MiffreEPJD06} and the Stark phase shift $\varphi_{S,i}$ for an
atom in the interferometer arm $i$ is given by:

\begin{eqnarray}
\label{hs3} \varphi_{S,i} & =& \frac{2 \pi \epsilon_0 \alpha}{\hbar
v} \int \mathbf{E}_i^2(z) dz \nonumber \\
&=&\frac{2 \pi \epsilon_0 \alpha}{\hbar v} \frac{V_i^2}{h_i^2} L_{i}
\end{eqnarray}

\noindent where $L_{i}= \left[2a_i - (2h_i/\pi)\right]$ is the
effective length of capacitor $i$ and $V_i$ the potential difference
across the capacitor. The small correction \cite{MiffreEPJD06} due to
the fact that the atom passes at a distance ca. $40$ $\mu$m of the
septum, is negligible. The Stark phase shift $\varphi_S$ is the
difference of these two phase shifts:

\begin{eqnarray}
\label{hs4} \varphi_S  & =& \varphi_{S,l} - \varphi_{S,u} = \frac{2
\pi \epsilon_0 \alpha}{\hbar v}\left[ \frac{V_l^2}{h_l^2} L_{l} -
\frac{V_u^2}{h_u^2} L_{u} \right]
\end{eqnarray}

\noindent where $l$ ($u$) refers to the upper (lower)
arm of the interferometer as schemed in fig. \ref{figure4new}. By
tuning the voltage ratio $V_u/V_l$, we can cancel $\varphi_S$ for all
atom velocities.

\subsection{Taking into account capacitor defects}
\label{s33}

The two capacitors present geometrical defects: the electrodes and
the septum are not perfectly plane and parallel and the design of the
guard electrodes is imperfect. We describe these defects by assuming
that the spacing $h_i(y,z)$ of capacitor $i$ is a slowly varying
function of $y$ and $z$ and  that the length $L_{i}(y)$ between guard
electrodes is a slowly varying function of $y$. Finally, the voltage
across the capacitor $i$ is the sum of the applied voltage $V_i$ and
of contact potentials $V_{c,i}(y,z)$ which is the difference of the
work functions of the two electrodes ($V_{c,i}(y,z)$ is of the order
of $100$ mV). An exact calculation of the field would be complicated
and we assume that the local field $E_i (y,z)$ is the field of a
perfect plane capacitor of spacing $h_i(y,z)$:

\begin{eqnarray}
\label{hs5}
E_i (y,z)& =&  \frac{V_i + V_{c,i}(y,z)}{h_i(y,z)}
\end{eqnarray}

\noindent The phase $\varphi_{S,i}$ is a function of $y$:

\begin{eqnarray}
\label{hs6}
\varphi_{S,i}(y)& =&  \frac{2 \pi \epsilon_0 \alpha}{\hbar v} \int_{L_{i}(y)} E_i^2 (y,z) dz
\end{eqnarray}

\noindent In an exact calculation of $E_i (y,z)$, the small-scale
variations of $h_i(y,z)$ and $V_{c,i}(y,z)$ would be washed out
because the atoms sample the electric field at a distance ca. $40$
$\mu$m of the septum and only variations with a scale larger than
this distance may play a role. We will not try to take this effect
into account but most of the rapid variations of the electric field
are already washed out in the phase because of the integral appearing
in equation (\ref{hs6}). The calculation of $\varphi_{S,i}(y)$ is
detailed in Appendix B. Because $V_{c,i}(y,z)$ is always
much smaller than $V_i$, the quadratic term in $V_{c,i}(y,z)$ is
negligible and we get $\varphi_{S,i}(y)
=\varphi_{S,g,i}(y)+\varphi_{S,c,i}(y) $ with a dominant term
$\varphi_{S,g,i}\propto V_i^2$  and a minor term
$\varphi_{S,c,i}(y)\propto V_i V_{c,i}(y,z)$. The $y$-variations of
$\varphi_{S,g}$ are due to geometrical defects:

\begin{eqnarray}
\label{e5} \varphi_{S,g,i}(y)& =&  \frac{2 \pi \epsilon_0
\alpha}{\hbar v} V_i^2\int_{L_{i}(y)} \frac{dz}{h_i^2(y,z)}
\end{eqnarray}

\noindent while the $y$-variations of $\varphi_{S,c,i}(y)$ are mostly
due to contact potentials $V_{c,i}(y,z)$:

\begin{eqnarray}
\label{e6} \varphi_{S,c,i}(y)& =&  \frac{4 \pi \epsilon_0
\alpha}{\hbar v} V_i\int_{L_{i}(y)} \frac{V_{c,i}(y,z)
dz}{h_i^2(y,z)}
\end{eqnarray}

\noindent As in eq. (\ref{hs4}), the Stark phase shift is the
difference of the  phases on the two arms $i=l,u$.

\section{Effects of the magnetic field on the interferometer signals}
\label{s4}

In this section, we recall the hyperfine-Zeeman Hamiltonian and we
discuss the Zeeman phase  shifts resulting from a
gradient of the magnetic field between the two interferometer arms.

\subsection{The hyperfine-Zeeman Hamiltonian}
\label{s41}

For lithium ground state, the hyperfine-Zeeman Hamiltonian
$H_{HFS+Z}$ is given by:

\begin{equation}
\label{z1} H_{HFS+Z} = A \mathbf{I}\cdot \mathbf{S} -g_S \mu_B
\mathbf{S}\cdot \mathbf{B} - g_I \mu_B \mathbf{I}\cdot
\mathbf{B}
\end{equation}

\noindent $\mathbf{S}$ and $\mathbf{I}$ are the electronic ($S=1/2$)
and nuclear ($I=3/2$) spins respectively. The ground state is split
in two hyperfine levels $F=1,2$ and 8 $F,m_F$ sublevels. The
Fermi-contact hyperfine parameter $A$, the electronic Land\'e factor
$g_S$ and the nuclear Land\'e factor $g_I$ are very accurately known
\cite{ArimondoRMP77,YanPRL01}.

We have omitted the diamagnetic term $H_{dia} = -\sum_i q^2
(r_{\perp,i}^2) \mathbf{B}^2/8m$, where $r_{\perp,i}$ is the
projection of the nucleus-electron vector on a plane perpendicular to
$\mathbf{B}$. Using $\sum_i<r_i^2>$ given by ref. \cite{YanPRA95},
for our largest field, $B_{max} \approx 1.4 \times
10^{-2}$ T, the energy shift is $\Delta E_{dia} \approx 2.4 \times
10^{-32}$ J, which is very small and independent of the sublevel.
Moreover, as the interferometer signal is sensitive only to the
difference of $\Delta E_{dia}$ between the two interferometer arms
and as the magnetic field homogeneity is very good, the resulting
phase is fully negligible.

\begin{figure}
\begin{center}
\includegraphics[width = 8 cm]{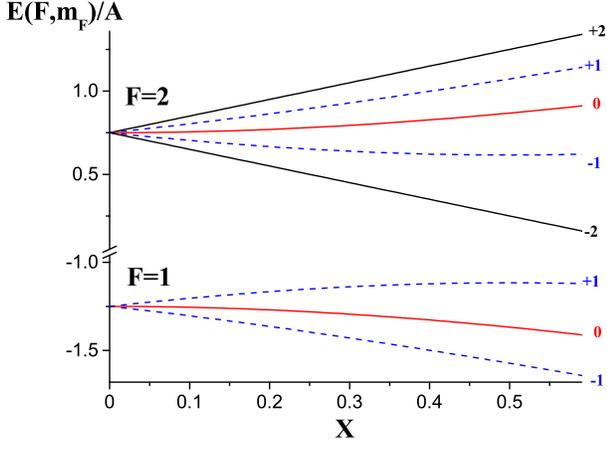}
\caption{(color online). Hyperfine-Zeeman energies $E(F,m_F)/A$ plotted as a function
of $X = - \left( g_S  - g_I \right)\mu_B B /(2A)$. It appears clearly
that there are four pair of levels with almost opposite Zeeman energy
shifts. \label{figure5new}}
\end{center}
\end{figure}

The Zeeman energy shifts are always smaller than $\mu_B B_{max}
\approx 1.3 \times10^{-25}$ J and the ratio of these shifts to the
kinetic energy is smaller than $2\times 10^{-5}$, which remains
small. The magnetic field extends over $\approx 80$ mm corresponding
to an interaction time $t_{int} \approx 75 $ $\mu$s. If the magnetic
field was applied to one interferometer arm only, a first-order
perturbation calculation predicts a maximum phase $\varphi_{Z,max} =
\mu_B B_{max}t_{int}/\hbar \approx 10^5$ rad and the second-order
term of the perturbation expansion is of the order of $1$ rad, which
is not at all negligible. In our experiment, the field homogeneity is
good, $\Delta B/B \approx 10^{-4}$, where $\Delta B$ is the
difference of the field on the two interferometer arms: with this
field difference, the first order term induces a Zeeman phase
shift of the order of $10$ rad at most, while the
second order terms compensate each other and their contribution to
the Zeeman phase shift is negligible, below $1$ mrad.
Finally, hyperfine uncoupling cannot be neglected for our maximum
field and the hyperfine Zeeman energies are given by:

\begin{eqnarray}
\label{z2} E(F,m_F, B) &=& -\frac{A}{4} - g_I \mu_B m_F B  \pm
A\sqrt{1 + m_F X + X^2} \nonumber \\
\mbox{with  }X &=& - \frac{\left( g_S  - g_I \right)\mu_B B }{2A}
\end{eqnarray}

\noindent  with $X = 34.9 B$ ($B$ in Tesla) so that for our largest
field $X\approx 0.5$. If $X<1$, the $\pm$ sign is associated to the
$F = I \pm 1/2$ level.  If we forget the small $g_I \mu_B m_F B$ term
in eq. (\ref{z2}), there are four pairs of levels with opposite
Zeeman energy shifts, the three pairs of levels with the same $m_F$
value and the pair $F=2, m_F = \pm 2$ and this property will be
useful. The variations of $E(F,m_F, X)$ are plotted in fig.
\ref{figure5new}. Later, we will use the derivatives of
$E(F, m_F, X)$ with respect to $X$, given by:

\begin{eqnarray}\label{z3}
\frac{\partial E(F, m_F, B)}{A \partial X} &\approx& - g_F m_F \pm
\left[1 -\frac{m_F^2}{4}\right]X \nonumber \\
&& \pm \frac{3m_F}{4}\left[\frac{m_F^2}{4}-1\right]X^2
\end{eqnarray}

\noindent This expansion, limited to the $X^2$ terms, is exact for
the $F=2, m_F = \pm2$ sublevels.  For $\left|X\right| <
0.5$, its accuracy is better than $3$\% for the $m_F=\pm 1$
sublevels, but only $12$\% for the $m_F=0$ sublevels.

\subsection{Calculation of the Zeeman phases and their effects on the fringe phase and visibility}
\label{s42}

If the magnetic field never vanishes and if its direction is slowly
varying along the atom trajectory, it is a good approximation to
assume an adiabatic behavior
\cite{SchmiedmayerJPII94,giltner95,JacqueyEPL07} : the projection
$m_F$ of the total angular momentum $\mathbf{F}$ on an axis parallel
to $\mathbf{B}$ is constant and the Zeeman phase is given by:

\begin{eqnarray}
\label{z6} \varphi_Z \left( F, m_F\right)  & =&  - \frac{1}{\hbar v} \oint_{l-u}
E(F, m_F,B) ds \nonumber \\
&\approx  & \frac{1}{\hbar v} \int \frac{\partial E(F,
m_F,B)}{\partial B} \frac{\partial B}{\partial x} \delta x(z)  dz
\end{eqnarray}

\noindent where $\delta x(z)$ is the distance between the
interferometer arms at the coordinate $z$ and $B$ is the modulus of
the magnetic field.  When the magnetic field is produced
by a current $I$ circulating in a coil, the dependence with $I$ of
the Zeeman phase shifts are complicated. We obtain an approximate
analytic expression using the power expansion, eq. (\ref{z3}):

\begin{eqnarray}
\label{z7}
\varphi_Z \left( F, m_F\right) & =&  -g_Fm_F J_1 \pm  \left[1 -\frac{m_F^2}{4}\right] J_2 \nonumber \\
&& \pm \frac{3m_F}{4}\left[\frac{m_F^2}{4}-1\right]J_3\nonumber \\
J_1 &=& \frac{\mu_B}{\hbar v} \int \frac{\partial B}{\partial x} \delta x(z)  dz \nonumber \\
J_2 &=& \frac{\left( g_S  - g_I \right)^2\mu_B^2}{8A \hbar v}\int \frac{\partial (B^2)}{\partial x} \delta x(z) dz\nonumber \\
J_3 &=& \frac{3\left( g_S  - g_I \right)^3\mu_B^3}{128A^2 \hbar v}\int \frac{\partial (B^3)}{\partial x} \delta x(z) dz
\end{eqnarray}

\noindent $J_k$ is proportional to ${\left|I\right|}^k$
and the Zeeman phase shifts are expressed as third order polynomials
of $I$. Moreover, the presence and inhomogeneity of the laboratory
field, which exists when $I=0$, must be taken into account. In this
aim, we introduce corrections to the linear Zeeman effect
(coefficient $J_1$): in consistency with the weak value of the
laboratory field, these corrections will be most accurate when the
field produced by the coil is weak.

\subsection{The case of linear Zeeman effect}

If the field $B$ is smaller than about $2\times 10^{-3}$ T
corresponding to $\left|X\right| <0.07$, Zeeman effect is almost
purely linear, with Land\'e factors $g_F$ equal to $g_{1} =
\left(-g_S + 5 g_I\right) /4 \approx -0.502 053 \approx -1/2$ and $
g_{2} = \left(g_S +3 g_I\right)/4 \approx 0.499 689 \approx 1/2$, the
approximate values $\pm 1/2$ being sufficiently accurate. Taking into
account the population unbalance described by the parameter $\chi$
given by eq. (\ref{app1}), the complex visibility defined by eq.
(\ref{a11}) and (\ref{a12}) is equal to:

\begin{eqnarray}
\label{z8}
\frac{\underline{\mathcal{V}}}{\mathcal{V}_0} &=& \frac{1}{4}  \left[1 + 2(1+ 5\chi)\cos\left(g_1J_1\right) + (1-3 \chi)\cos\left(g_2J_1\right)
\right.\nonumber \\ && \left. + (1-3 \chi)\cos\left(2g_2J_1\right)\right]
\end{eqnarray}

\begin{figure}
\begin{center}
\includegraphics[width = 8 cm]{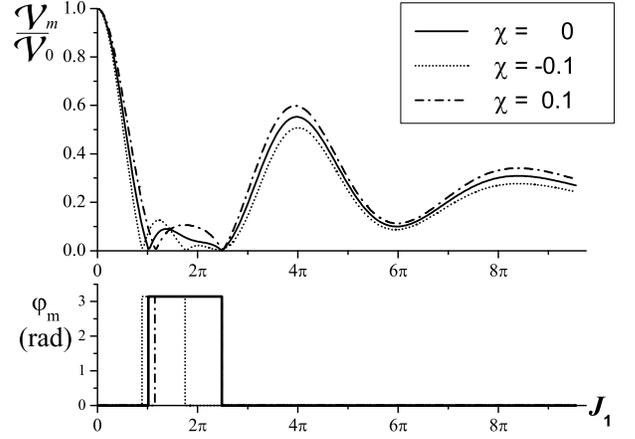}
\caption{Calculated relative visibility and phase as a function of
$J_1$, for a velocity distribution with a typical value of the
parallel speed ratio $S_{\|}= 8$ and for three values of the
population unbalance parameter $\chi$: $\chi=0$ full curve;
$\chi=0.1$ dash-dotted curve; $\chi= -0.1$ dotted curve. This
parameter has a large effect especially when the visibility is very
low. \label{figure6new}}
\end{center}
\end{figure}

\noindent In this case, the complex visibility remains real i.e. the
fringe phase is exactly equal to $0$ or $\pi$. For a well defined
atom velocity, when $J_1$ increases, the visibility first decreases
and presents revivals with $\mathcal{V} = \mathcal{V}_0$ when $J_1
/(4 \pi)$ is equal to an integer. In fig. \ref{figure6new}, the
modulus and the phase of the complex visibility are plotted as a
function of $J_1$, for different values of the parameter $\chi$, with
the velocity distribution parameter $S_{\|}=8$: the visibility
revivals are less intense because of the velocity average.

We now calculate corrections of $J_1$ to describe the
influence of the laboratory field $\mathbf{B}_0$, which is not
perfectly homogeneous. We express the total magnetic
field $\mathbf{B}_{tot} = \mathbf{B}_{coil}+ \mathbf{B}_0$, where
$\mathbf{B}_{coil}$ is the coil field proportional to the coil
current $I_{coil}$. Its modulus is:

\begin{eqnarray}
\label{z10} B_{tot} &=& \sqrt{\left(\mathbf{B}_{coil}+
\mathbf{B}_0\right)^2} \nonumber \\
&\approx & B_{coil}\left(1 \pm \frac{\mathbf{u}\cdot
\mathbf{B}_0}{B_{coil}}\right)
\end{eqnarray}

\noindent an approximation valid when $B_0 \ll B_{coil}$.
$\mathbf{u}$ is a local vector parallel to $\mathbf{B}_{coil}$ so
that $\mathbf{B}_{coil}= \pm \mathbf{u}B_{coil}$ where the $\pm$ sign
is the sign of  $I_{coil}$. We split the integral giving
$J_1$ in two regions, the region where the coil field is dominant
($in$) and the region where the laboratory field is dominant
($out$):

\begin{eqnarray}
\label{z11} J_1 &\approx& \frac{\mu_B}{\hbar v} \left[ \int_{in}
\frac{\partial B_{coil}}{\partial x} \delta x(z)  dz \pm \int_{in}
\frac{\partial \mathbf{u}\cdot \mathbf{B}_0}{\partial x} \delta x(z)
dz \right.\nonumber \\ &&\left. + \int_{out} \frac{\partial
B_0}{\partial x} \delta x(z)  dz \right]
\end{eqnarray}

\noindent The first term, proportional to $\left|I_{coil}\right|$,
is written $A_{J1,coil} \left|I_{coil}\right|$. The
second term is constant and it is convenient to write it
$-A_{J1,coil} I_{0,coil}$ which defines a quantity $I_{0,coil}$
homogeneous to a current. The third term is independent of the
current in the coil and we write it $J_{0,coil}$. In this way, we
get:

\begin{eqnarray}
\label{z12}
J_1 = A_{J1,coil} \left|I_{coil}-I_{0,coil} \right| + J_{0,coil}
\end{eqnarray}

\noindent It is important to note that $J_{0,coil}$ depends on the
coil because of integration in the $out$ region. We call $J_0$ the
integral as in equation (\ref{z11}) extended to the whole
interferometer:

\begin{eqnarray}
\label{z13} J_0 &=& \frac{\mu_B}{\hbar v} \int \frac{\partial
B_0}{\partial x} \delta x(z)  dz
\end{eqnarray}

\noindent We need a formula which interpolates smoothly when
$I_{coil}$ varies. When $I_{coil} \rightarrow 0$, the quantity $J_1$
must tend toward $J_0$. This property is verified by eq. (\ref{z12})
if we take $J_{0,coil}= J_{0} - A_{J1,coil} \left|I_{0,coil}
\right|$. Finally, as we use two coils, a main coil
(current $I$) and a compensator coil (current $I_C$), we generalize
eq. (\ref{z12}) which becomes:

\begin{eqnarray}
\label{z14} J_1 =  A_{J1} \left|I-I_{0} \right|+ A_{J1,C}
\left|I_{C}-I_{0,C} \right| + J_{0,I+C}
\end{eqnarray}

\noindent where $J_{0,I+C} = J_0 - A_{J1} \left| I_0 \right| - A_{J1}
\left| I_{0,C} \right|$. To establish eq. (\ref{z14}), we must assume that the ($in$)
regions of the two coils do not overlap, which is satisfied by our
experimental apparatus \cite{LepoutrePhD,LepoutreXXX}. Eq. (\ref{z14}) will be
used to fit experimental data.

\subsection{The case of larger magnetic fields}

We consider here only $J_1$ and $J_2$ to simplify the equations. The
complex visibility is then given by:

\begin{eqnarray}
\label{z15} \frac{\underline{\mathcal{V}}}{\mathcal{V}_0} &=&
\frac{1}{4} \left[(1+ \chi)\left(\cos\left(J_2\right)+  2
\cos\left(\frac{3J_2}{4}\right) \cos\left(\frac{J_1}{2}\right)
\right) \right.\nonumber \\
&& \left. + (1-3 \chi)\cos\left(J_1\right) \right]\nonumber \\  && +
i \chi \left[\sin\left(J_2\right) + 2
\cos\left(\frac{J_1}{2}\right)\sin\left(\frac{3J_2}{4}\right)
\right]\nonumber \\
\end{eqnarray}

\begin{figure}
\begin{center}
\includegraphics[width = 8 cm]{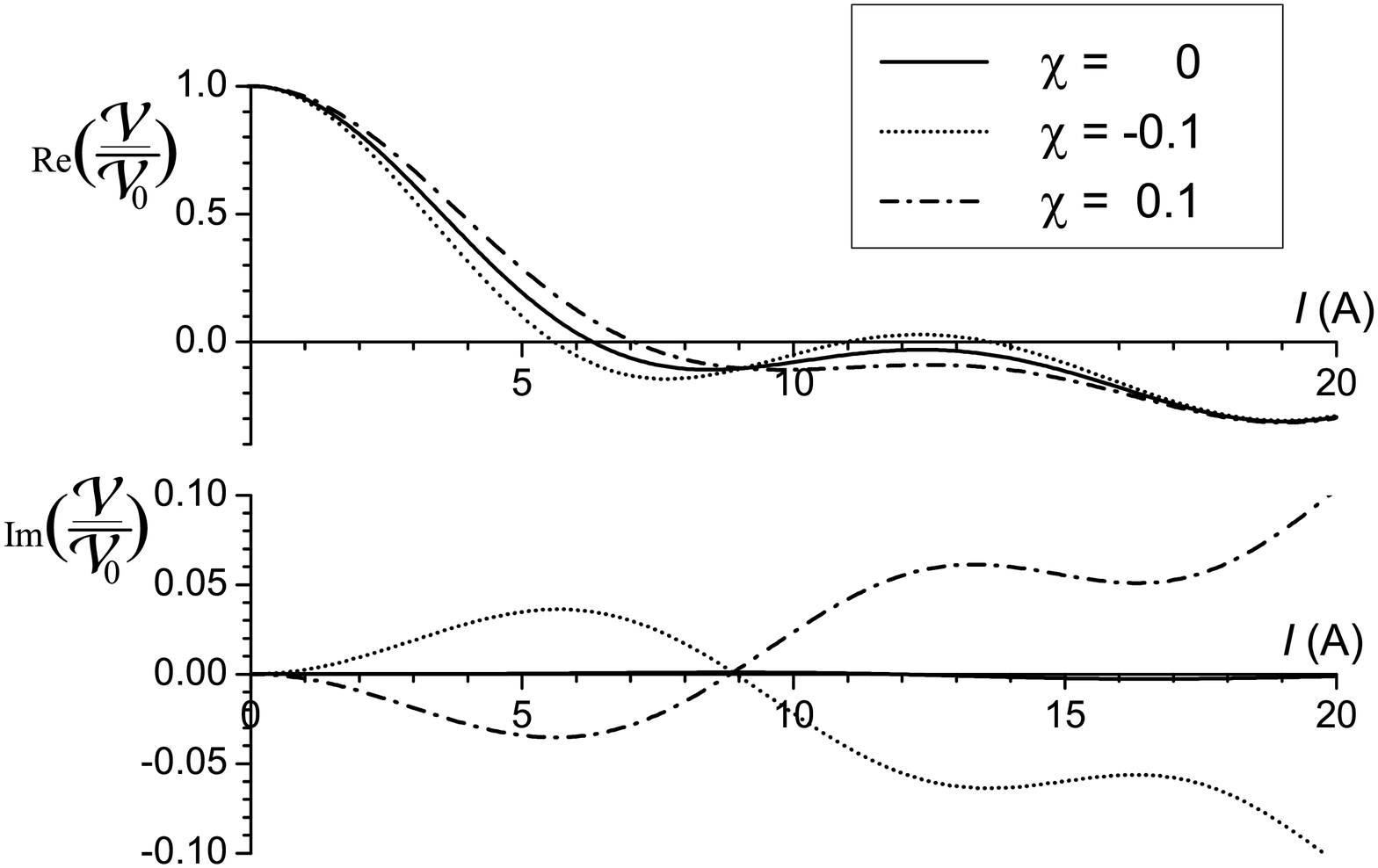}
\caption{Calculated real and imaginary parts of the complex
visibility $\underline{\mathcal{V}}/\mathcal{V}_0$ as a function of
the coil current $I$. We use the values $J_1/I= 0.5$ rad/A and
$J_2/I^2= 0.01$ rad/A$^2$, which are close to our experimental
values. Same $S_{\|}$ and $\chi$ values as in fig.
\ref{figure6new}.\label{figure7new}}
\end{center}
\end{figure}

\noindent In fig. \ref{figure7new}, we have plotted the complex
fringe visibility as a function of the magnetic field inhomogeneity.
When $\chi =0$, the imaginary part of $\underline{\mathcal{V}}$
almost vanishes but it differs slightly from $0$ because we have
taken into account the nuclear spin contribution (an effect neglected
in eq. (\ref{z15})).  When $\chi$ differs from $0$, the imaginary
part is not at all negligible  and the fringe phase may be large, of
the order of $1$ rad, when the real part of the visibility is small.

We should also calculate corrections to $J_2$ and $J_3$ for the
inhomogeneity of the laboratory field, but these
refinements are expected to be of weak influence and did not appear
to improve the quality of the fits. As a consequence, only the
correction to $J_1$ given by eq. (\ref{z14}) have been taken into
account.

\section{The Aharonov-Casher phase shift}
\label{s5}

As explained above, the Aharonov-Casher phase $\varphi_{AC}$, given by eq. (\ref{PhiAC}),
can be considered as being due to the motional magnetic field $\mathbf{B}_{mot} = \mathbf{ E}
\times \mathbf{ v}/c^2$. This field is usually very small, $B_{mot}
\approx 10^{-8}$ T for our largest electric field $ E_{max} \approx
0.8$ MV/m and $v=v_m = 1065$ m/s, but it has opposite
values on the two interferometer arms as we use opposite electric
fields. In practice, $B_{mot}$ is always smaller than $10^{-3}$ of
the magnetic field and only the component of $B_{mot}$ parallel to
this local magnetic field can play a role, but it cannot
be neglected. The magnetic moment $\mu (F,m_F)$ of the $F,m_F$
sublevel is a function of the magnetic field $\mathbf{B}$.
We introduce a vector $\mathbf{u}_{tot} = \mathbf{B}/B$
parallel to the total magnetic field at the location $\mathbf{r}$, and we approximate the magnetic dipole moment value by
using an expansion similar to eq. (\ref{z3}):

\begin{eqnarray}
\label{f13}
\boldsymbol{\mu} (F,m_F)&=& \mu(F,m_F)  \mathbf{u}_{tot} \nonumber \\
\mbox{with  } \mu(F,m_F)&=& \mp \mu_B \frac{m_F + 2 X}{2\sqrt{1 + m_F X+ X^2}}\nonumber \\
&\approx& \mp \mu_B \left[\frac{m_F}{2} + \left(1 -\frac{m_F^2}{4}\right)X \right. \nonumber \\
&+& \left.  \frac{3m_F}{4}\left(\frac{m_F^2}{4}-1\right)X^2 \right]
\end{eqnarray}

\noindent with the notations of eq.  (\ref{z2}).  We then use equation (\ref{PhiAC}) to calculate the AC phase shift as a function of $F$, $m_F$.

\section{Summary of the various phase shifts}
\label{s6}

In this section, we rapidly review the various phase shifts discussed in the previous sections and we estimate their magnitude in our experimental setup. We also explain their effects on the fringe visibility. The phase $\varphi_p$ in equation (\ref{a0}) is the sum of 5 contributions:

\begin{eqnarray}
\label{sum1}
\varphi_p &=& \varphi_{Sagnac} + \varphi_{S}  + \varphi_Z \left( F, m_F\right) \nonumber \\&&+ \varphi_{AC}(F,m_F) + \varphi_{HMW}
\end{eqnarray}

\begin{table}[b]
 \begin{tabular}{|l|c|c|c|c|c|}
    \hline
    Phase Shift   &  Maximum value  &     Dependence   & Effect on         \\
                  &  (rad)          &  with $F$, $m_F$ & fringe visibility \\ \hline
    Sagnac        & $0.64$          &        no        & negligible        \\ \hline
    Polarizability& $\approx 0.1$   &        no        & weak              \\ \hline
    Zeeman        & $\approx10$     &       yes        & strong            \\ \hline
    AC            & $0.07$          &       yes        & weak              \\ \hline
    HMW           & $0.027$         &       no         & no                \\ \hline
\end{tabular}
\caption{The phase shifts present in our experiment: for each phase shift, we give its value for the maximum fields $E_{max}\approx  0.8$ MV/m and $B_{max}\approx 14 $ mT available in the interaction region used for the detection of the HMW phase, the existence of a dependence with the sublevel and its effect on the fringe visibility.   \label{table1}}
\end{table}

\noindent Let us discuss each term separately:

\begin{itemize}

\item the Sagnac phase shift  $\varphi_{Sagnac}$ due to Earth rotation is easily calculated from the latitude of our experiment and the size of the interferometer:

\begin{equation}
\label{sum2}
\varphi_{Sagnac}=688/v
\end{equation}
\noindent where $v$ is the atom velocity in m/s and $\varphi_{Sagnac}$ is measured in rad. With $v_m = 1065$ m/s, this phase is rather small, $\varphi_{Sagnac}\approx 0.65$ rad \cite{JacqueyPRA08} and, as its dispersion is solely due to its velocity dependence, it has only minor effects on the fringe visibility $\mathcal{V}$.

\item the Stark phase shift $\varphi_{S}$ can be very large, about $300$ rad if we applied the largest electric field  $E= 0.8$ MV/m on one arm only. Because of its velocity dependence, $\varphi_{S} \propto 1/v$, the fringe visibility $\mathcal{V}$ decreases when $\varphi_{S}$ increases and becomes very small for $\varphi_{S}> 30$ rad because the velocity distribution of our atomic beam has a relative full width of the order of $25$\%. In order to measure the HMW phase, we need the best possible fringe visibility and we tune the electric fields on the two arms so that the mean $\varphi_{S}$  is of the order of $100$ mrad. The reduction of fringe visibility due to the velocity averaging is then completely negligible but, because of defects of the  geometry of the two capacitors, the $y$-dependence of $\varphi_{S}(y)$ discussed above induces a minor reduction of the fringe visibility.

\item  the Zeeman phase shift $ \varphi_Z \left( F,m_F\right)$ would be extremely large, about $10^5$ rad if our maximum field $B= 14 $ mT was applied on one interferometer arm only, but with a relative field difference $\delta B/B \sim 10^{-4}$ between the two interferometer arms, the Zeeman phase shift is reduced to about $10$ rad for the $F=2, m_F=\pm2$ sublevels. Because of the dependence of $ \varphi_Z$ with $F,m_F$ and with the atom velocity, $\varphi_{Z} \propto 1/v^2$, this phase shift would still be sufficient to reduce the fringe visibility to a very small value. A compensating coil creating an opposite field gradient between the two interferometer arms is necessary to preserve a good visibility but, because of non-linear Zeeman effect due hyperfine uncoupling, this compensation is not complete.

\item the Aharonov-Casher phase shift $\varphi_{AC}(F,m_F)$ is a function of the $F$, $m_F$ sublevel and it is largest for the $F=2$, $m_F= \pm 2$ sublevels. Because of its geometric character, it is independent of the atom velocity. For our largest electric field, $\varphi_{AC}(F=2, m_F= 2)\approx 70 $ mrad. Because of its $F, m_F$-dependence, the AC phase shift has a weak but detectable effect on the fringe visibility.

\item the He-McKellar-Wilkens phase shift  $\varphi_{HMW}$ is independent of the $F,m_F$ hyperfine sublevel and of the atom velocity, because of its geometric character. For our largest electric and magnetic fields, $\varphi_{HMW} \approx 27 $ mrad. As the HMW phase shift is not dispersed, it has no effect on the fringe visibility.

\end{itemize}

Table \ref{table1} summarizes the main properties of these phase shifts present in our experiment. We have two comments. The existence of  phase shifts larger than the one we want to measure is not a problem as long these large phase shifts are stable: in order to observe the weak HMW phase shift, we subtract the phase shift due to the electric field and the one due to the magnetic field from the one observed when both fields are applied. The real problem comes from the fact that the signal is the sum of the signals due to 8 hyperfine sublevels and, as shown by equation (\ref{a13}), the weights of the sublevel $j$ is the product $P_j  \mathcal{V}_j$. The visibility $\mathcal{V}_j$ varies with the applied perturbations and this is the basis of systematic effects analyzed in section \ref{s2}.

\section{Conclusion}

In this paper, we have recalled what are the topological phases of
electromagnetic origin, namely the Aharonov-Bohm, the Aharonov-Casher
and He-McKellar-Wilkens phases and the theoretical connections
between these various effects. We have also discussed the possible
detection schemes of the HMW phase and we have explained the
principle of our experiment based on a separated-arm lithium-atom
interferometer.

During our experiment, which is briefly described in ref.
\cite{LepoutrePRL12} (with more details in the companion paper HMWII
\cite{LepoutreXXX}), we have observed unexpected stray phases: most
of them have been explained by our calculations and they result from
phase-averaging effects due to experimental defects. We have
discussed these effects on general grounds in section \ref{s2}.

In order to develop a model of our experiment, we have analyzed in
detail the Stark and Zeeman effective Hamiltonian in the
$^2$S$_{1/2}$ ground state of $^7$Li atom and we have discussed the
validity of several approximations. We have thus shown that we may
assume that the Stark shift is independent of the $F,m_F$ sublevel
and that the diamagnetic term is negligible. We have explained why we
use a first-order calculation of the Stark and Zeeman phases. We have
also discussed in detail the phase shifts resulting of the
inhomogeneities of the electric or magnetic fields and the
consequences of these phase shifts on the fringe phase and
visibility. Finally, we have evaluated the Aharonov-Casher phase in
our experiment and shown that it is small but not fully negligible.

\acknowledgments

We thank CNRS INP, ANR (grants ANR-05-BLAN-0094 and
ANR-11-BS04-016-01 HIPATI) and R\'egion Midi-Pyr\'en\'ees for
supporting our research.

\section{Appendix A: Relative contributions of the $F,m_F$ sublevels to the signal}

In this appendix, we discuss various effects which may modify the
relative populations of the $F, m_F$ sublevels.

\subsection{The populations of the $F, m_F$ sublevels in the incident atomic beam}

The atomic beam, when it enters the atom interferometer, is not
optically pumped. We may assume that the $8$ Zeeman-hyperfine
sublevels are equally populated for the following reasons:  the only
effects which could induce a partial selection of the internal states
are the supersonic expansion and Stern-Gerlach forces and they are
too weak to play a role in our experiment.

Supersonic expansions are well known to align the rotational angular
momentum of molecules by collisions with the carrier gas, because the
collisions between the seeded molecule and the carrier gas are not
isotropically distributed, an anisotropy due to the velocity
difference between the two species (the so-called velocity slip
effect). A similar effect can align an atomic angular momentum.
However, lithium atom in its ground state is in a spin $1/2$ state
which cannot be aligned. Nuclear spins are uncoupled during a
collision, because of the weakness of the hyperfine Hamiltonian with
respect to a typical collision duration, below $10^{-12}$ s, so that
collisions are not expected to align the total angular momentum
$\mathbf{F}$.

Stern-Gerlach forces due to a magnetic field gradient can deflect
differently the various $F, m_F$ sublevels and a  $F,m_F$-dependent
deflection can produce a population unbalance between these
sublevels. The only places where such a deflection could occur are in
the collimation slits and this would require a magnetic field
gradient of the order of $10^3$ Tesla/m. We have chosen to use
collimation slits made of silicon, a non-ferromagnetic material, so
that the magnetic field gradient is surely very small.

\subsection{The transmission of the interferometer}

As we are using linear polarization of the laser standing waves and
as the hyperfine structure of the $^2$P first resonance state of
lithium is quite small, the diffraction amplitude is independent of
$m_F$ for a given $F$ level \cite{ChampenoisPhD}. This is true even
in the presence of a weak magnetic field, comparable to the Earth
field, $4\times 10^{-5}$ T, because the Zeeman splitting of the
transitions, of the order of $1$ MHz in frequency units, is
negligible with respect to the laser frequency detuning
$\delta_L/(2\pi) \sim 2$ GHz \cite{MiffreEPJD05}.

\begin{figure}
\begin{center}
\includegraphics[height = 5 cm]{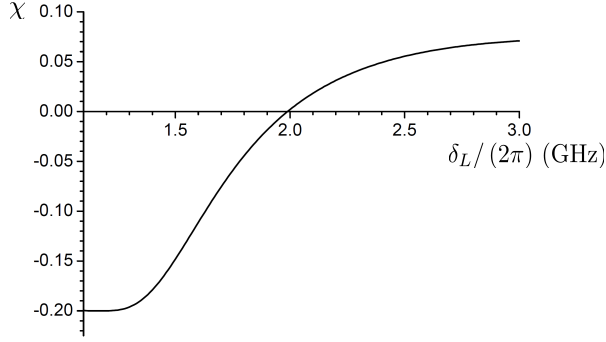}
\caption{The population unbalance parameter $\chi$ as a function of
laser frequency detuning $\delta_L/(2\pi)$ GHz as given by the simple
model (eq. \ref{app3}). The $\beta$ parameter has been taken equal to
$\beta/(2\pi) = 3.65$ GHz, which is optimum for a detuning
$\delta_L/(2\pi)= 2 $ GHz. \label{figure8new}}
\end{center}
\end{figure}

The diffraction amplitude still depends on $F$ because the laser
frequency detuning is not the same for the two hyperfine levels, the
ground state hyperfine splitting being equal to $\omega_{HFS}/(2\pi)=
0.803 $ GHz in frequency units. We define the population unbalance by
writing the relative population $P(F,m_F)$ of the $F,m_F$ sub-level
in the form:
\begin{eqnarray}
\label{app1}
P(F,m_F) &=& \left(1+ 5 \chi \right)/8 \mbox{   if  } F=1 \nonumber \\
P(F,m_F) &=& \left(1- 3 \chi \right)/8 \mbox{   if  } F=2
\end{eqnarray}
\noindent The total population is normalized and the unbalance
parameter $\chi$ must verify $-1/5 \leq  \chi \leq 1/3$ so that
$P(F,m_F)\geq 0$. We have developed a simplified model of the
interferometer transmission, neglecting the variation of the
diffraction amplitudes with the atom velocity vector (the modulus of
the velocity has a distribution given by eq. \ref{a2}, with $S_{\|}
\approx 8$ and the direction of the velocity vector is characterized
by an angular distribution with a full width at half maximum close to
$20$ $\mu$rad). In this way, we can write the first-order diffraction
amplitude by the $i^{th}$ laser standing wave in the form:

\begin{eqnarray}
\label{app2}
\left|\alpha_i\right| &=& \sin\left(\beta_i/\delta_L(F)\right)\nonumber \\
\mbox{with  } \delta_L(F=1) &=& \delta_L \nonumber \\
\mbox{and  } \delta_L(F=2) &=& \delta_L + \omega_{HFS}
\end{eqnarray}

\noindent where $\beta_i$  is a parameter proportional to the
integral of the laser power density seen by an atom which crosses the
$i^{th}$ laser standing wave. We assume that the $\beta_i$ parameters
are optimum for a Mach Zehnder interferometer with $\beta_2 =
2\beta_1 = 2\beta_3 =\beta $. The transmission of the interferometer
is proportional to $\left|\alpha_i\right|^4$ and we thus get the
unbalance parameter $\chi$:

\begin{eqnarray}
\label{app3}
\chi =  \frac{\sin^4\left(\beta/\delta_L\right) - \sin^4\left(\beta/\left(\delta_L+ \omega_{HFS}\right)\right)}
{3\sin^4\left(\beta/\delta_L\right) +5 \sin^4\left(\beta/\left(\delta_L+ \omega_{HFS}\right)\right)}
\end{eqnarray}

\noindent The variations of the unbalance parameter $\chi$ are
plotted as a function of the laser detuning $\delta_L$ in fig.
\ref{figure8new}, for a typical value of the experimental
parameter $\beta$.

\section{Appendix B: the Stark phase including capacitor defects}

The Stark phase $\varphi_i(y)$, given by eq. (\ref{hs6}), is
proportional to the integral $\int E_i^2 (y,z) dz$. We use an overline $
\bar{...}$ to note the average over $z$ defined by the integral over the
capacitor length $L_{i}(y)$, for instance, $\bar{h}_i(y) =   \int
h_i(y,z) dz/ L_i(y)$ and we note $\delta_i(y,z)$ the dimensionless
deviation from a plane capacitor defined by $h_i(y,z) = \bar{h}_i(y)
\left[1 + \delta_i(y,z)\right]$. By definition, $\bar{\delta}_i(y) =
0$ and we assume that $\delta_i(y,z)\ll 1$. We also define
$\bar{V}_{c,i}(y)=  \int V_{c,i}(y,z) dz/L_i(y)$.

To calculate the Stark phase $\varphi_i(y)$, we expand the electric
field $E_i (y,z)$ up to first order in  $\delta_i(y,z)$ and in
$V_{c,i}(y,z)/V_i$. Both assumptions are excellent, first
because the design of the capacitors ensures $\delta_i \ll 1$,
secondly because the contact potential term is of the order of $\pm
100$ mV while the applied voltage $V_i$ is of the order of $100$ V at
least (when $ V_i=0$, the Stark phase solely due to contact
potentials, of the order of $10^{-5}$ rad at most, is completely
negligible). Then $E_i^2 (y,z)$ is given by:

\begin{eqnarray}
\label{e8} E_i^2(y,z)& \approx&  \frac{V_i^2
}{\left(\bar{h}_i(y)\right)^2} \left[ 1 + 2 \frac{V_{c,i}(y,z)}{V_i}
-2 \delta_i(y,z)\right]
\end{eqnarray}

\noindent The phase $\varphi_i(y)$ is obtained by integration over
$z$:

\begin{eqnarray}
\label{e9} \varphi_i(y) & \approx & \frac{2 \pi \epsilon_0 \alpha
V_i^2}{\hbar v} \frac{L_i(y) }{ \left(\bar{h}_i(y)\right)^2} \left[ 1
+ 2 \frac{\bar{V}_{c,i}(y)}{V_i} \right]
\end{eqnarray}

\noindent We introduce $\eta_i(y)$ which  measures the $y$-dependence
of the $z$-integrated geometrical defect of the capacitor $i$.
$\eta_i(y)$ measures the relative $y$-variation of the $z$-averaged
thickness of the capacitor $i$; it is defined by:

\begin{eqnarray}
\label{e10} \frac{L_i(y) }{ \left(\bar{h}_i(y)\right)^2} = \left<
\frac{ L_i }{ \bar{h}_i^2} \right>  \left[1+ \eta_i(y)\right]
\end{eqnarray}

\noindent where $\left< ... \right>$ denotes the $y$-average with
the weight function $P(y)$. By definition, $\left<\eta_i \right> =
0$. We get:

\begin{eqnarray}
\label{e11} \varphi_i(y) & \approx &  \varphi_{0i} \left[1+
\eta_i(y)+ 2 \frac{\bar{V}_{c,i}(y)}{V_i} \right] \nonumber \\
\mbox{with  } \varphi_{0i}& =& \frac{2 \pi \epsilon_0 \alpha
V_i^2}{\hbar v} \left< \frac{ L_i }{ \bar{h}_i^2} \right>
\end{eqnarray}

\noindent In the HMW detection experiments discussed in HMWII
\cite{LepoutreXXX}, the voltage ratio $V_l/V_u$ is tuned so that it compensates the fact that the two capacitors have not exactly the same value of the quantity $\left< L_i / \bar{h}_i^2 \right>$. In this way we get
$\left|\left<\varphi_{l}\right>/ \left<\varphi_{u}\right> - 1 \right|
< 10^{-3}$. Hence for the defect terms which are
expressed by a first order expansion, it is justified to use the
mean value $\varphi_{0}$ of the induced phases $\varphi_{0i}$
and  the mean value  $V$ of the voltages $V_i$. We thus obtain the Stark phase shift, including the influence of the
capacitor defects:

\begin{eqnarray}
\label{e12} \varphi_S (y) &=& \varphi_{0l} - \varphi_{0u} +
\left<\varphi_{Sc}\right> \nonumber \\
&&+ \delta \varphi_{Sg}(y) + \delta \varphi_{Sc}(y)
\end{eqnarray}

\noindent In eq. (\ref{e12}), the mean phase shift (first line) is
given by:

\begin{eqnarray}
\label{e13} \left<\varphi_S\right> & = &  \varphi_{0l}-\varphi_{0u} +
\left<\varphi_{Sc} \right> \nonumber \\
\mbox{with  } \left<\varphi_{Sc} \right>  &=& 2 \varphi_{0} \left[
\frac{\left<\bar{V}_{c,l}\right> -\left<\bar{V}_{c,u}\right>}{V}
\right]
\end{eqnarray}

\noindent The term $\left(\varphi_{0l}-\varphi_{0u}\right)$, which is
dominant if the voltage ratio $V_u/V_l$ is not perfectly tuned, scales like $V^2$. The mean term due to the contact
potentials $\left<\varphi_{Sc}\right>$, which is expected to be
considerably smaller, scales like $V$. The dispersion of the Stark
phase shift with the atom trajectory is described by the terms of the
type $\delta \varphi(y)$ (second line of eq. \ref{e12}), with
$\left<\delta\varphi\right>=0$. The dispersion due to geometrical
defects scale like $V^2$, while that due to the contact potentials
scale with $V$:

\begin{eqnarray}
\label{e14} \delta \varphi_{Sg}(y) & = & \varphi_{0} \left[\eta_l(y)-
\eta_u(y)\right] \nonumber \\
\delta \varphi_{Sc}(y) & = & 2 \varphi_{0}
\frac{\bar{V}_{c,l}(y)-\bar{V}_{c,u}(y)}{V} - \left< \varphi_{Sc}
\right>
\end{eqnarray}

\noindent Although we expect the dispersion originating from the
contact potentials to be smaller, and to exhibit weak correlations
because of rapid small-scale variations, its influence could not be
ruled out prior to our experiment.


\end{document}